\documentclass[%
 aip,
 sd,%
 amsmath,amssymb,
reprint,%
]{revtex4-1}

\usepackage{multirow}
\usepackage{graphicx}
\usepackage{bm}% bold math
\usepackage{hyperref}% add hypertext capabilities
\usepackage{xcolor}

\usepackage{graphicx, textcomp, amssymb, mathrsfs}

\makeatletter
\newcommand\xleftrightarrow[2][]{%
\ext@arrow 9999{\longleftrightarrowfill@}{#1}{#2}}
\newcommand\longleftrightarrowfill@{%
\arrowfill@\leftarrow\relbar\rightarrow}
\makeatother

\usepackage{epstopdf}

\begin{document}

\title{Procedure for Obtaining the Analytical Distribution Function of Relaxation Times  for the Analysis of Impedance Spectra using the Fox $H$-function}

\author{Anis Allagui$^*$}
\email{aallagui@sharjah.ac.ae}
\affiliation{Dept. of Sustainable and Renewable Energy Engineering, University of Sharjah, Sharjah, P.O. Box 27272, United Arab Emirates}
\altaffiliation[Also at ]{Center for Advanced Materials Research, Research Institute of Sciences and Engineering, University of Sharjah, Sharjah, P.O. Box 27272,  United Arab Emirates}
\affiliation{Dept. of Electrical and Computer Engineering, Florida International University, Miami, FL33174, United States}

\author{Ahmed Elwakil} 
%\email{hbenaoum@sharjah.ac.ae}
\affiliation{
Dept. of Electrical Engineering, 
University of Sharjah, PO Box 27272, Sharjah, United Arab Emirates 
}

\begin{abstract}

The interpretation of electrochemical impedance spectroscopy data by fitting it to equivalent circuit models has been a standard method of analysis in electrochemistry. 
%However, selecting a specific circuit requires not only enough prior knowledge about the system, but also it is often found that many similar circuits can fit the data equally well. 
However, the  inversion of the data from the frequency domain to a distribution function of relaxation times (DFRT) has gained considerable attention for impedance data analysis, as it can  reveal more detailed information about the underlying electrochemical processes without requiring a priori knowledge. 
The focus of this paper is to provide a general procedure for obtaining analytically the DFRT  from an impedance model, assuming an elemental Debye relaxation model as the kernel. 
%The DFRT is very useful for analyzing and understanding electrochemical impedance spectroscopy  data. 
The procedure consists of first representing the impedance function in terms of the Fox $H$-function, which possesses many useful properties particularly that its Laplace transform is again an  $H$-function. From there the DFRT is obtained by two successive iterations of inverse Laplace transforms. In the passage, one can easily obtain an expression for the   response function to a step excitation. The procedure is tested and verified on some known impedance models. 
   
%\begin{tocentry}
%\includegraphics[width=5cm]{toc.pdf} 
%\end{tocentry}


\end{abstract}

\maketitle

\section{Introduction}

%\subsection{Background}

Electrochemical impedance spectroscopy (EIS) is one of the principal techniques used in electrochemistry.  EIS data are acquired by applying a small amplitude voltage $v(t)$ (or current $i(t)$) perturbation at different frequencies to an electrochemical system, and measuring the resulting current (or voltage). 
Applying small magnitude perturbations is meant to suppress nonlinear behaviors, allowing the system to be studied using relatively simple linear models. 
The impedance (or admittance) spectrum is then defined as the point-by-point ratio between the voltage and the current, both expressed in the frequency domain, i.e. $V(s)$ and $I(s)$ ($s=i\omega$). Note that $V(s)$ and $I(s)$ are the Laplace transforms (LT) of $v(t)$ and $i(t)$, knowing that the  LT of a function $f(t)\;(t\in\mathbb{R}^+)$ is defined by 
%\begin{equation}
$\mathcal{L}[f(t);s] = F(s)= \int_0^{\infty} e^{-st} f(t) dt,\;s\in\mathbb{C}$
%\end{equation}
 and the inverse LT of $F(s)$  is defined by 
%\begin{equation}
$\mathcal{L}^{-1}[F(s);t] = f(t) = {(2\pi i)^{-1}} \int_{\gamma-i\infty}^{\gamma+i\infty} e^{st} F(s) ds,\;\mathcal{R}e(s)=\gamma$.  
 % \end{equation} 
The impedance of a system is thus the complex-valued function \cite{barsoukov2018impedance, lasia2014electrochemical}:
\begin{equation}
Z(f) = \frac{V(s)}{I(s)} = Z'(s) + j Z"(s)
\label{eq:Z}
\end{equation}
which can be used, in principle, to characterize and extract valuable physical, chemical and electrical information about the system under study (e.g.  charge transport and transfer processes, charge storage, aging and degradation \cite{barsoukov2018impedance,lasia2014electrochemical} 
 in batteries \cite{single2019theory, bredar2020electrochemical, vadhva2021electrochemical}, supercapacitors \cite{eis},  fuel cells \cite{he2009exploring}, solar cells \cite{guerrero2021impedance}, electrochemical sensors \cite{pejcic2006impedance}, corroding surfaces \cite{mansfeld1990electrochemical}, biomaterials \cite{CESEWSKI2020112214, lisdat2008use}, etc.). In this regard, the success of EIS as a useful technique is contingent on the ability of one to describe the system via a quantitative physical model, but this may  not often be possible  \cite{effendy2020analysis}. 

 %Despite the widespread use of EIS, obtaining meaningful insights from the data can be difficult. 
The interpretation of EIS data is commonly done via its  (complex nonlinear least squares) fitting  to  electrical models consisting of  resistors, capacitors, inductors, fractional capacitors and fractional inductors arranged in series and/or parallel combinations \cite{bredar2020electrochemical}. However, this may not always represent the actual physics of the system because the selection of an equivalent circuit is based on the user's experience and  prior understanding of the system \cite{ciucci2015analysis}. Furthermore, multiple equivalent circuits can actually fit the same data equally well. The data fitting problem can be seen differently if Eq.\,\ref{eq:Z},   is rewritten in the time domain with the use of the convolution theorem as \cite{allagui2021inverse, ieeeted}:
\begin{equation}
v(t) = (z \circledast i) (t) 
= \int\limits_{-\infty}^{+\infty} z(\xi) i(t-\xi)  d\xi 
\end{equation}
The time-domain function $z(t)$, being the inverse LT of $Z(s)$, represents the collective electrical response to whatever changes are happening from within the system of physical and chemical nature. 

If $z(t)$ is considered to be an average relaxation function such that when a current impulse is subjected onto the system the resulting voltage will relax monotonically to zero, we can write $z(t)$ as \cite{ciucci2015analysis}:
\begin{equation}
z_D(t) = c_0 \delta(t) + \sum\limits_{n=1}^N {c_n} \theta(t) \tau_n^{-1} {e^{-t/\tau_n}}
\label{eq:z}
\end{equation}
where $\theta(t)$ is the Heaviside theta function, equal to $0$ for $t<0$ and $1$ for $t>0$, $c_n\;(n=0,1,\ldots,N)$ are constant coefficients, and $\tau_n$ are positive time constants. The term $c_0 \delta(t)$ represents the response of the system at $t=0$. 
%A time constant is the time that a system takes in returning to the steady state after a perturbation is applied. 
It is understood that Eq.\,\ref{eq:z} is obtained by considering the system under test to be of capacitive nature, and comprised of an infinite series of $RC$ elements according to the Debye circuit model. We consider from now on that the impedance function is normalized with respect to an arbitrary resistance $R$.

By taking the LT of Eq.\,\ref{eq:z} we obtain the impedance of the system in the frequency domain as:
\begin{equation}
Z_D(s)= c_0 + \sum\limits_{n=1}^N \frac{c_n}{1+ s \tau_n}
\label{eq:Zf}
\end{equation}
With $c_0=R_0/R$ being a normalized ohmic resistance, $c_n=g_n \Delta \tau$ ($\Delta \tau \to 0$), $N\to\infty$, we rewrite the impedance in Eq.\;\ref{eq:Zf} as:
\begin{equation}
Z_D(s) \approx R_0/R + \int\limits_{0}^{\infty} \frac{g(\tau)}{1+ s \tau} d\tau
\label{eq:Zf2}
\end{equation}
or, as usually preferred with $\gamma(\ln\tau) = \tau g(\tau)$, as
\begin{equation}
Z_D(s) \approx R_0/R + \int\limits_{0}^{\infty} \frac{\gamma(\ln\tau)}{1+ s \tau} d\ln\tau
\label{eq:Zf22}
\end{equation}
Here in $Z_D(s)$ the frequency is the independent variable, while $R_0$ and the distribution function of relaxation times (DFRT)  $g(\tau)$ are the unknown model parameters \cite{effendy2020analysis}. 
We note that the time-domain function $z_D(t)$ in Eq.\;\ref{eq:z} can also be expressed as the integral of exponential decays as follows:
\begin{equation}
z_D(t) = (R_0/R)\, \delta(t) + \int\limits_{0}^{\infty}  {g(\tau)} \theta(t) {\tau^{-1}} e^{-t/\tau} d\tau
\label{eq:z2}
\end{equation}

The elementary Debye kernel in Eq.\ref{eq:Zf2} 
representing an exponential relaxation process 
  can be replaced with other functions.  For example, Florsch, Revil, and Camerlynck \cite{florsch2014inversion} considered the Havrilliak-Negami (HN) model as a kernel. This gives by superposition the following expression for the impedance function:
\begin{equation}
Z_{HN}(s) \approx R_0 + \int\limits_{0}^{\infty} \frac{g(\tau)}{(1+ (s \tau)^{\alpha})^{\beta}} d\tau \quad ( 0< \alpha, \beta  \leqslant 1)
\label{eq:ZfHN}
\end{equation}
in which the problem is now to retrieve $g(\tau)$ assuming $\alpha$ and $\beta$ are known a priori. 
It is clear that for  $\beta=1$ we recover the Cole-Cole model, with $\alpha=1$ the  Davidson-Cole model is recovered, and with  $\alpha=\beta=1$ we retrieve back the Debye model.

The aim of this work is to provide a general procedure for recovering analytically the function $g(\tau)$ from Eq.\;\ref{eq:Zf2} with the use of Fox's $H$-function and its properties (Section\;\ref{sec:method}).  
We note that finding the distribution of the relaxation times can be  directly obtained from the experimentally-measured spectral data (i.e. without the need to model it) via  numerical inversion methods, such as by Fourier transform techniques \cite{schichlein2002deconvolution, boukamp2015fourier}, Tikhonov regularization \cite{gavrilyuk2017use, paul2021computation, shanbhag2020relaxation}, Max Entropy \cite{horlin1998deconvolution, horlin1993maximum}, genetic algorithm \cite{tesler2010analyzing}, etc. \cite{boukamp2020distribution, effendy2020analysis, liu2020deep, ciucci2015analysis}. However, it is also instructive to have the analytical expression for the DFRT  associated with a given impedance model.  
 For this purpose, we shall first recall from the general treatment of Macdonald and Brachman\;\cite{macdonald1956linear} on integral transforms in linear systems some of the useful relations to be used for this analysis (Section\;\ref{sec:basic}). The authors provided there a comprehensive set of relations between the various complex functions used to describe networks and systems, as well as between responses to various types of inputs. These functions can be for instance an impedance or an admittance, a transfer ratio, a complex susceptance, etc. \cite{macdonald1956linear}.
 Next, after presenting the proposed method of inversion from Eq.\;\ref{eq:Zf2} with the use of the $H$-function (Section\;\ref{sec:method}), we derive  analytical expressions pertinent to systems described with some of the most widely used models of impedance (Section\;\ref{sec:examples}).% with the use of the $H$-function and its relevant properties. 

\section{Theory}
\subsection{Basic relations}
 \label{sec:basic}
 
\begin{figure}[t]
\begin{center}
\includegraphics[width=0.39\columnwidth,angle=-90]{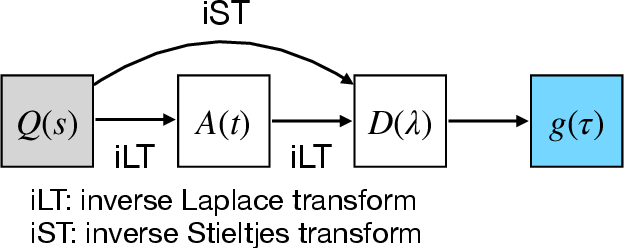}
\caption{Basic relations between the functions $Q(s)$, $A(t)$, $D(\lambda)$ and $g(\tau)$ ($g(\tau)= \tau^{-1} D(\tau^{-1})$ with $\tau= \lambda^{-1}$)}
\label{fig0}
\end{center}
\end{figure}

 Using the definitions and notations of Macdonald and Brachman, we define from Eq.\;\ref{eq:Zf2} the integral transform:
 \begin{equation}
Q(s) = \int\limits_{0}^{\infty} \frac{g(\tau)}{1+ s \tau} d\tau
\label{eq:Zf02}
\end{equation} 
as the reduced impedance function for the simple Debye dispersion model $(1+ s \tau)^{-1}$.  
If we define an inverse  time constant $\lambda=\tau^{-1}$, and a distribution $D(\lambda)$ as
\begin{equation}
D (\lambda) = 
\lambda^{-1} g(\lambda^{-1}) 
\end{equation}
 then $Q(s)$ in Eq.\;\ref{eq:Zf02} becomes the iterated LT of $D(\lambda)$ as\;\cite{macdonald1956linear}:
 \begin{equation}
Q(s) = \int\limits_{0}^{\infty} \frac{D (\lambda) }{ s + \lambda} d\lambda = \mathcal{L} \{ \mathcal{L}[D(\lambda);t];s \}
\label{eq:LTLT}
\end{equation}
This can be readily deducted from:
\begin{equation}
Q(s) = \int\limits_{0}^{\infty} {D (\lambda) }  d\lambda  \int\limits_{0}^{\infty}   e^{- t (s+\lambda)}  dt\, 
\end{equation} 
We recognize that this iterated LT is the Stieltjes transform (ST) \cite{bateman1954tables}. 
 The system response $A(t)$ to a step function  can be obtained from:
\begin{equation}
A(t) = \mathcal{L}^{-1}[Q(s);t]
\label{eq:At}
\end{equation} 
With Eq.\;\ref{eq:LTLT}, we can  also write
\begin{equation}
A(t) = \mathcal{L}[D(\lambda);t] %= \int\limits_0^{\infty} \lambda^{-1} g(\lambda^{-1}) e^{-t \lambda} d \lambda
\end{equation}
By inverse LT we have:
\begin{equation}
D(\lambda) = \mathcal{L}^{-1}[A(t);\lambda)] %=\lambda^{-1} g(\lambda^{-1}) 
\end{equation}
and we can obtain $g(\tau)$ from:
\begin{equation}
g(\tau) = \tau^{-1} D(\tau^{-1})
\label{Eq:g}
\end{equation}
Fig.\;\ref{fig0} shows schematically the connection between the different functions $Q(s)$, $A(t)$, $D(\lambda)$ and $g(\tau)$.

One simple example to demonstrate these steps is to consider  the case of a series $RC$ circuit with an 
 admittance   given by
 \begin{equation}
Y(s) = \frac{sC}{1+RCs}
\end{equation}
In normalized form this admittance is:
\begin{equation}
\frac{Y(s)}{sC} = 
Q_1(s) = \frac{1}{1+s\tau_0}
\end{equation}
where $\tau_0=RC$ is a time constant. 
Its inverse LT gives:
\begin{equation}
A_1(t) = {\tau_0}^{-1} e^{-t/\tau_0}
\end{equation}
and by another iteration of inverse LT we obtain:
\begin{equation}
D_1(\lambda) = \tau_0^{-1}\delta(\lambda - \tau_0^{-1})
\end{equation}
Using Eq.\;\ref{Eq:g}, we obtain for the DFRT:
\begin{equation}
g_1(\tau) = \tau^{-1} \tau_0^{-1}\delta( \tau^{-1} - \tau_0^{-1}) = \delta(\tau-\tau_0)
\end{equation}
which is as expected an impulse delayed by $\tau_0$. 
%More useful examples are given in Macdonald and Brachman\;\cite{macdonald1956linear}.

%We note at this point that one can apply the Titchmarsh inversion formula (\# 11.8.4 in \cite{titchmarsh1948introduction}) directly on Eq.\;\ref{eq:LTLT} to recover $D(\lambda)$, and then $g(\tau)$. This is done by putting $s=e^{\xi}$, $\lambda=e^{\eta}$, $\psi(\xi)=e^{\xi/2} Q(e^{\xi})$, $\phi(\eta)=e^{\eta/2} D(e^{\eta})$ which leads to:
%\begin{equation}
%\psi(\xi) = \int\limits_{-\infty}^{\infty}  \frac{1}{2} \phi(\eta) \, \text{sech}\left(\frac{\xi-\eta}{2}\right)  d\eta
%\end{equation} 
%Recall that the Fourier transform  
%(FT) of a function $f(t)$ is defined by 
% $\mathcal{F}[f(x);u] = F(u)= {(2\pi)^{-1/2}} \int_{-\infty}^{\infty} e^{i u x} f(x) dx$
%  and the inverse FT of $F(u)$  is defined by 
% $\mathcal{F}^{-1}[F(u);x] = f(x) = {(2\pi)^{-1/2}} \int_{-\infty}^{\infty} e^{-i x u} F(u) du$.  
% Now using the convolution theorem of the FT we obtain:
%\begin{align}
%\phi(\xi) &= ({\pi \sqrt{2\pi}})^{-1} \int\limits_{-\infty}^{\infty} \Psi(u) \cosh(\pi u) e^{-i\xi u} du \\
%&= {(2\pi)^{-3/2}} \int\limits_{-\infty}^{\infty}  \Psi(u)  \left(e^{-i (\xi + i\pi) u}   + e^{-i (\xi - i\pi) u} \right) du \\
%&= {(2\pi)^{-1}}  \left[ \psi(\xi + i\pi) + \psi(\xi - i\pi) \right]
%\end{align}
%or
%\begin{equation}
%D(\lambda)=\frac{i}{2\pi} 
%\left[ Q(x e^{ i\pi}) - Q(x e^{- i\pi}) \right]
%\label{eq:iS} 
%\end{equation}

The procedure for deriving expressions for $g(\tau)$ can be streamlined and generalized with  use of Fox's $H$-function \cite{fox1961g}, as we show below. The  use of the  $H$-function is motivated by the fact that it has many convenient properties  and amongst them is that its Laplace transform is again an  $H$-function \cite{mathai2009h}. This allows us to obtain both the system response function and the DFRT in a straightforward manner in terms of the $H$-function. 
%In the following sections of the manuscript we derive  analytical expressions pertinent to systems described with some of the most widely used models of impedance. 

\subsection{Proposed method of inversion}
\label{sec:method}
The procedure we propose here for deconvolving the DFRT $g(\tau)$ from Eq.\;\ref{eq:Zf2} requires first to express a given impedance function in terms of Fox's $H$-function. 
We recall that Fox's $H$-function  of order 
$(m,n,p,q)\in \mathbb{N}^4$, ($0 \leqslant n \leqslant p$, $1 \leqslant m \leqslant q$) 
and with parameters 
  $A_j \in \mathbb{R}_+ \;(j=1,\ldots,p)$, $B_j \in \mathbb{R}_+\; (j=1,\ldots,q) $, 
$a_j \in\mathbb{C}\;(j=1,\ldots,p)$ and $ b_j \in \mathbb{C}\; (j=1,\ldots,q)$ 
is defined for $z \in \mathbb{C},\;z\neq 0$ by the contour  integral  \cite{fox1961g, mathai2009h}:
\begin{equation}
H^{m,n}_{p,q}\left[ z|^{(a_1,A_1),\ldots,(a_p,A_p)}_{(b_1,B_1),\ldots, (b_q,B_q)} \right] =\frac{1}{2\pi i} \int_L h(s) z^{-s} ds
\label{eq:H}
\end{equation}
 where the integrand $h(s)$ is given by:
 \begin{equation}
h(s) = \frac{\left\{\prod\limits_{j=1}^m \Gamma(b_j + B_j s)\right\}  \left\{\prod\limits_{j=1}^n \Gamma(1-a_j - A_j s)\right\}}
{\left\{\prod\limits_{j={m+1}}^q \Gamma(1-b_j - B_j s)\right\} \left\{\prod\limits_{j={n+1}}^p \Gamma(a_j + A_j s)\right\}}
\end{equation}
In Eq.\;\ref{eq:H}, $z^{-s}=\exp \left[ -s (\ln|z|+ i \arg z) \right] $ and $\arg z$ is not necessarily the principal value. 
 The contour of integration $L$ is a suitable contour separating the poles  
 of  $\Gamma(b_j+ B_j s)$ ($j=1,\ldots,m$) from the poles 
 of   $\Gamma (1-a_{j} - A_{j} s)$ ($j=1,\ldots,n$).
 An empty product is always interpreted as unity. 
 For more details on the $H$-function including definition, convergence, and many useful properties we refer to  Mathai,  Saxena, and   Haubold \cite{mathai2009h}, Mathai and  Saxena \cite{mathai1978h}, and Kilbas and Saigo \cite{kilbas2004h}.  
 
The importance of the  $H$-function for this work arises from that fact that (i) it contains a vast number of elementary and special functions used in science and engineering as special cases; see for instance Appendices A.6 and A.7 in Mathai and  Saxena \cite{mathai1978h} for many examples of elementary functions expressed in terms of $H$-functions, and $H$-functions expressed in terms of elementary functions. Furthermore, (ii) the  Laplace transform of an $H$-function is again an  $H$-function, which is essentially the main mathematical tool needed to deal with Eq.\;\ref{eq:LTLT}. This result is obtained from the following steps \cite{hilfer, glockle1993fox}:
\begin{align}
\mathcal{L} 
&\left[ 
 H_{p,q}^{m,n}\left[ t\left|
\begin{array}{c}
(a_p,A_p)  \\
(b_q,B_q)  \\
\end{array}
\right.\right]; u
\right] \\ \nonumber
&= 
\frac{1}{2\pi i} \int_0^{\infty} \int_L e^{-u t} h(s) t^{-s}  ds \,dt
 \\ \nonumber 
 & = \frac{1}{2\pi i}  \int_L h(s) u^{s-1} \Gamma(1-s) ds   \\ \nonumber
 &= 
  u^{-1} 
H^{m,n+1}_{p+1,q}\left[   u^{-1}\left|
\begin{array}{c}
(0,1), (a_1,A_1), \ldots, (a_p,A_p)   \\
(b_1,B_1), \ldots, (b_q,B_q)  \\
\end{array}
\right.\right]
\end{align}
Furthermore, we have the inverse LT given by \cite{glockle1993fox}:
\begin{align}
\mathcal{L}^{-1} 
&\left[ 
 H^{m,n}_{p,q}\left[  u\left|
\begin{array}{c}
(a_p,A_p)  \\
(b_q,B_q)  \\
\end{array}
\right.\right];t
\right] \\ \nonumber
 &= 
 t^{-1} 
H^{m,n}_{p+1,q}\left[  t^{-1} \left|
\begin{array}{c}
(a_p,A_p), \ldots, (a_1,A_1), (0,1)   \\
(b_1,B_1), \ldots, (b_q,B_q)  \\
\end{array}
\right.\right]
\end{align} 
From Mathai,  Saxena and Haubold \cite{mathai2009h} (formula 2.19), we   have:
\begin{align}
\mathcal{L} 
&\left[ 
t^{\rho-1} H_{p,q}^{m,n}\left[ a t^{\sigma} \left|
\begin{array}{c}
(a_p,A_p)  \\
(b_q,B_q)  \\
\end{array}
\right.\right];  u
\right] \\ \nonumber
 &= 
  u^{-\rho} 
H^{m,n+1}_{p+1,q}\left[ a  u^{-\sigma}\left|
\begin{array}{c}
(1-\rho,\sigma), (a_1,A_1), \ldots, (a_p,A_p)   \\
(b_1,B_1), \ldots, (b_q,B_q)  \\
\end{array}
\right.\right]
\end{align}
which uses extra parameters ($a,\rho,\sigma$) in case they are  needed. Its inverse LT  (formula 2.21 in \cite{mathai2009h}) is:
\begin{align}
\mathcal{L}^{-1} 
&\left[ 
  u^{-\rho} 
H_{p,q}^{m,n}\left[ a  u^{\sigma}\left|
\begin{array}{c}
(a_p,A_p)  \\
(b_q,B_q)  \\
\end{array}
\right.\right]; t
\right] \label{eq:ILT} \\ \nonumber
 &= t^{\rho-1} 
H_{p+1,q}^{m,n}\left[ a t^{-\sigma}\left|
\begin{array}{c}
(a_p,A_p), \ldots, (a_1,A_1), (\rho,\sigma)  \\
(b_1,B_1),\ldots,(b_q,B_q)  \\
\end{array}
\right.\right]
\end{align}

Now using Eqs.\;\ref{eq:LTLT} and\;\ref{Eq:g} with the help of a few general properties of the $H$-function, mainly (i) identities dealing with the reciprocal of an argument \cite{kilbas2004h,mathai1978h,mathai2009h}:
\begin{equation}\label{H-inverse}
   H^{m,n}_{p,q}\left[ z 
   \left| 
  \begin{array}{l} 
  (a_p,A_p) \\
  (b_q,B_q) \\
  \end{array}
  \right. \right]
  = 
  H^{n,m}_{q,p}\left[ z^{-1} 
  \left| 
  \begin{array}{l}
  (1-b_q,B_q) \\
  (1-a_p,A_p) \\
  \end{array} 
  \right. \right],
\end{equation} 
and (ii) the multiplication of an $H$-function by the argument to a certain power:
\begin{equation}\label{H-power}
  z^\sigma H^{m,n}_{p,q}\left[ z 
  \left| 
  \begin{array}{l} 
  (a_p,A_p) \\
  (b_q,B_q) \\
  \end{array}
  \right.
  \right]
  = H^{m,n}_{p,q}\left[ z 
  \left| 
  \begin{array}{l} 
  (a_p+\sigma A_p,A_p) \\
  (b_q+\sigma B_q,B_q) \\
  \end{array} 
   \right.
  \right] 
\end{equation}
leads to the desired result for the DFRT $g(\tau)$.   
 In Section\;\ref{sec:examples} below we apply this procedure to a few examples of well-known impedance functions.
 
\section{Examples}
\label{sec:examples}

\subsection{Constant phase element}
 The constant phase element (CPE) is widely used in equivalent circuit models for impedance fitting of anomalous data that cannot be easily described with basic $R$ and $C$ circuit elements \cite{eis, lasia2022origin, gateman2022use, foedlc, allagui2021inverse}. 
 Its impedance function is given by:
 \begin{equation}
Z_c(s) = \frac{1}{C_{\alpha} s^{\alpha}}
\label{eq:CPE0} 
\end{equation}
where $C_{\alpha}$ is a pseudo-capacitance in units of F\,s$^{\alpha-1}$  and $\alpha$ is known as the dispersion coefficient. For $0<\alpha \leqslant 1$, the CPE represents the impedance of a fractional capacitor of constant phase $\phi(Z_c)=\tan^{-1}(-\alpha \pi/2)$, and for $\alpha=1$, it represents the impedance of an ideal capacitor.  For the particular case of $\alpha=0.5$, it represents the Warburg impedance.
 Normalizing the impedance function in Eq.\;\ref{eq:CPE0} 
  to an arbitrary resistance $R$  gives:
\begin{equation}
Q_{c}(s) = \frac{1}{(s\tau_{c})^{\alpha}}
\label{eq:CPE} 
\end{equation}
where $\tau_{c} = (RC_{\alpha})^{1/\alpha}$ is a characteristic time constant.    
Applying two successive times the inverse LT to $Q_{c}(s)$   gives first the system response function as:
\begin{equation}
A_{c}(t)
%=\mathcal{L}^{-1}[Q_{c}(s);t] 
= \frac{(t/\tau_{c})^{\alpha}}{ t\, \Gamma(\alpha)}
\end{equation}
and then the distribution $D_{c}(\lambda)$ as:
\begin{equation}
D_{c}(\lambda) 
%&= \mathcal{L}^{-1}[A_{c}(t);\lambda)]  \\
=  \pi^{-1} \sin(\alpha\pi)  ( \tau_{c}\,\lambda ) ^{-\alpha} 
\label{eq:dclambda}
\end{equation}
We remind that $  {\pi}/{\sin(\pi z)} = \Gamma(z) \Gamma(1-z) 
$ 
 (reflection formula for the gamma function). 
%\begin{equation}
%\Gamma(z) \Gamma(1-z) = \frac{\pi}{\sin(\pi z)}
%\end{equation} 
Using Eq.\;\ref{Eq:g} we obtain the DFRT $g_{c}(\tau)$ as the power-law function \cite{barsoukov2018impedance}:
\begin{equation}
g_{c}(\tau) = \pi^{-1} \sin(\alpha\pi) \tau_{c}^{-\alpha} \tau^{\alpha-1} 
\end{equation}
We note that $D_{c}(\lambda)$ can be obtained directly through the inverse ST applied onto Eq.\;\ref{eq:CPE}  using \cite{bateman1954tables}:
\begin{equation}
s^{-\alpha} = \frac{1}{\Gamma(\alpha) \Gamma(1-\alpha)} \int\limits_0^{\infty} \frac{\lambda^{-\alpha}}{s+\lambda} d\lambda  
\label{eq:salpha}
\end{equation}
 or thought the Titchmarsh inversion formula (\# 11.8.4 in \cite{titchmarsh1948introduction}):
\begin{equation}
D_c(\lambda)=\frac{i}{2\pi} 
\left[ Q_c(\lambda e^{ i\pi}) - Q_c(\lambda e^{- i\pi}) \right]
\label{eq:iS} 
\end{equation}
 Discretization of Eq.\;\ref{eq:salpha} has been adapted by Abdelaty et al. \cite{abdelaty2018approximation} as a way of  representing the $ s^{\alpha}$ term  as a weighted sum of first-order high-pass  filters.

Now in order to apply the proposed inversion method, we rewrite the CPE impedance function in Eq.\;\ref{eq:CPE} as an $H$-function 
 using the following relation\;\cite{mathai2009h}:
\begin{align}
{z^{\gamma}}(1- z)^{\beta} =  \Gamma(\beta+1) 
H^{1,0}_{1,1}\left[  z  \left|
\begin{array}{c}
(\gamma+\beta+1,1)    \\
(\gamma,1) \hfill   \\
\end{array}
\right.\right]
\end{align} 
which gives:
\begin{align}
\label{eq:QsCPE}
Q_{c}(s) =   
H^{1,0}_{1,1}\left[  (s\tau_{c})  \left|
\begin{array}{c}
(-\alpha+1,1)    \\
(-\alpha,1)  \hfill \\
\end{array}
\right.\right]
\end{align}
 valid for the argument $|s\tau_{c}|<1$, and
\begin{align}
\label{eq:QsCPE1}
Q_{c}(s) =   
H^{0,1}_{1,1}\left[  (s\tau_{c})  \left|
\begin{array}{c}
(-\alpha+1,1)    \\
(-\alpha,1)  \hfill \\
\end{array}
\right.\right]
\end{align}
valid for $|s\tau_{c}|>1$. 
We note that when $A_i=B_j=1$ $(i=1,\ldots,p)$; $(j=1,\ldots,q)$ the $H$-function reduces to the Meijer $G$-function \cite{kilbas2004h}. 
This makes Eq.\;\ref{eq:QsCPE1} for instance to be: 
\begin{align}
Q_{c}(s) =   
G^{0,1}_{1,1}\left[  (s\tau_{c})  \left|
\begin{array}{c}
(-\alpha+1)    \\
(-\alpha)  \hfill \\
\end{array}
\right.\right]
\end{align} 
The inverse LT of Eq.\;\ref{eq:QsCPE1} using the formula given in \ref{eq:ILT} (with the help of the identity\;\ref{H-inverse}) is:
\begin{align}
A_{c}(t)& =   
t^{-1} H^{0,1}_{2,1}\left[  (\tau_{c}/t)  \left|
\begin{array}{c}
(-\alpha+1,1),(0,1)    \\
(-\alpha,1)  \hfill \\
\end{array}
\right.\right] \\
&=   
t^{-1} H^{1,0}_{1,2}\left[  (t/\tau_{c})  \left|
\begin{array}{c}
(1+\alpha,1)  \hfill \\
(\alpha,1),(1,1)    \\
\end{array}
\right.\right] \\
&= \frac{(t/\tau_{c})^{\alpha}}{ t\, \Gamma(\alpha)}
\end{align} 
{from which:}
\begin{align}
D_{c}(\lambda)=   
 H^{1,0}_{2,2}\left[  (\lambda\, \tau_{c} )^{-1}  \left|
\begin{array}{c}
(1+\alpha,1),(1,1)    \\
(\alpha,1) ,(1,1) \hfill \\
\end{array}
\right.\right] 
\end{align} 
Then using Eq.\;\ref{Eq:g}, we obtain the DFRT for a CPE impedance as:
\begin{align}
g_{c}(\tau)= 
\tau^{-1}   
 H^{1,0}_{2,2}\left[  (\tau/\tau_{c} )  \left|
\begin{array}{c}
(1+\alpha,1) ,(1,1)  \\
(\alpha,1),(1,1)   \hfill \\
\end{array}
\right.\right]
%\tau_{c}^{-1}   
% H^{0,1}_{2,2}\left[  (\tau/\tau_{c} )  \left|
%\begin{array}{c}
%(\alpha,1) ,(0,1) \hfill \\
%(\alpha-1,1),(0,1)    \\
%\end{array}
%\right.\right] 
\end{align} 

\begin{figure}[!t]
\begin{center}
\includegraphics[width=0.98\columnwidth]{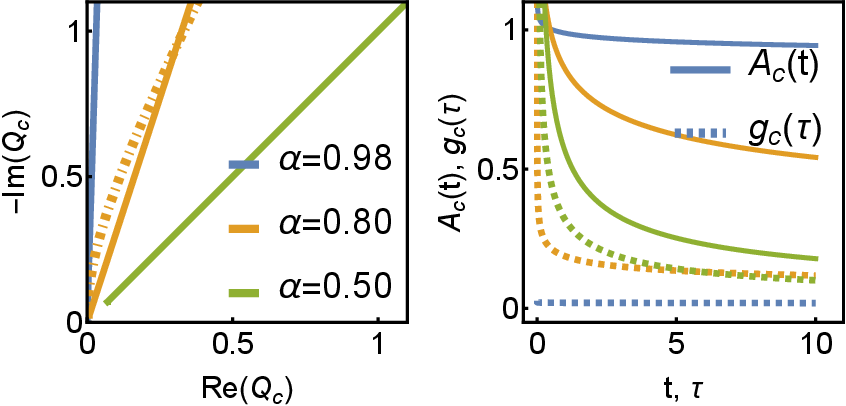}
\caption{Plots of (a) $Q_c(s)$ (Eqs.\;\ref{eq:QsCPE} and\;\ref{eq:QsCPE1}) in Nyquist form of real vs. imaginary parts for $\tau_c=1$ and  $\alpha=$ 0.98, 0.80 and 0.50. In dot-dashed we plot the discretized version of Eq.\;\ref{eq:Zf02} with $g(\tau)=g_c(\tau)$ for values of $\tau$ varying from 0.5 to 1000\,s at a step of 0.1\,s. In (b) we plot the functions $A_c(t)$ and $g_c(\tau)$ for the same parameters:  $\tau_c=1$ and  $\alpha=$ 0.98, 0.80 and 0.50}
\label{fig1}
\end{center}
\end{figure}

In Fig.\;\ref{fig1} we show plots of the three functions $Q_c(s)$, $A_c(t)$ and $g_c(\tau)$ for the case of
$\tau_c=1$ and for the three values of $\alpha=0.98, 0.80, 0.50$ . The frequency range for the Nyquist plot of $Q_c(s)$ is 0.01 to 100\,Hz. One can clearly see the gradual deviation of the plots from the ideal case of a capacitor ($\alpha=1$) as the value of $\alpha$ is reduced. 
We also plotted in Fig.\;\ref{fig1}(a), in dot-dashed line, the discretized version of Eq.\;\ref{eq:Zf02} with $g(\tau)=g_c(\tau)$ for $\alpha=0.80$ and for values of $\tau$ varying from 0.5 to 1000\,s at a step of 0.1\,s.

\subsection{Davidson-Cole Model}

The Davidson-Cole model is given by \cite{davidson1951dielectric}:
\begin{equation}
Q_{\gamma}(s) = \frac{1}{(1+s\tau_{\gamma})^{\gamma}} \quad (0< \gamma \leqslant 1)
\label{eq:QsalphaCD}
\end{equation}
Using the formula\;\cite{mathai2009h}:
\begin{align}
(1- z)^{-\alpha} =  \frac{1}{\Gamma(\alpha)} 
H^{1,1}_{1,1}\left[ - z  \left|
\begin{array}{c}
(1-\alpha,1)    \\
(0,1) \hfill   \\
\end{array}
\right.\right]
\end{align} 
we rewrite $Q_{\gamma}(s)$ as:
\begin{align}
Q_{\gamma}(s) =  \frac{1}{\Gamma(\gamma)} 
H^{1,1}_{1,1}\left[ s\tau_{\gamma}  \left|
\begin{array}{c}
(1-\gamma,1)    \\
(0,1) \hfill   \\
\end{array}
\right.\right]
\label{eq:QgammaH}
\end{align} 
The inverse LT of Eq.\;\ref{eq:QsalphaCD} is:
\begin{equation}
A_{\gamma}(t)=\frac{\left({t}/{\tau_{\gamma} }\right)^{\gamma } e^{-{t}/{\tau_{\gamma} }} }{t\, \Gamma (\gamma )}
\label{eq:Agammat0}
\end{equation}
which can be represented in terms of the $H$-function as:
\begin{equation}
A_{\gamma}(t)= \frac{1}{{\tau_{\gamma}\, \Gamma (\gamma )}}  
H_{0,1}^{1,0}\left[ t / \tau_{\gamma} \left|
\begin{array}{c}
-   \\
(\gamma-1,1) \hfill   \\
\end{array}
\right.\right]
\label{eq:Agamma}
\end{equation}
Using the inverse LT formula\;\ref{eq:ILT} on Eq.\;\ref{eq:QgammaH} gives:
\begin{align}
A_{\gamma}(t) 
&=
\frac{1}{t\,\Gamma(\gamma)} 
H^{1,1}_{1,2}\left[ t/\tau_{\gamma}  \left|
\begin{array}{c}
(1,1)  \hfill  \\
(\gamma,1),(1,1)    \\
\end{array}
\right.\right]
\label{eq:Agammat20}
 \\
&=  \frac{1}{\tau_{\gamma}\Gamma(\gamma)} 
H^{1,1}_{1,2}\left[ t/\tau_{\gamma}  \left|
\begin{array}{c}
(0,1)  \hfill  \\
(\gamma-1,1),(0,1)    \\
\end{array}
\right.\right]
\label{eq:Agammat2}
\end{align} 
The latter  can be readily reduced  to Eq.\;\ref{eq:Agamma} using\;\cite{mathai2009h}:
\begin{align}\label{H-reduction}
   H^{m,n}_{p,q}&  \left[ z 
   \left| 
  \begin{array}{l} 
  (a_1,A_1), \ldots, (a_p,A_p) \\
  (b_1,B_1), \ldots, (b_{q-1},B_{q-1}), (a_1,A_1) \\
  \end{array}
  \right. \right]
   \nonumber \\ 
   =&  
  H^{m,n-1}_{p-1,q-1}\left[ z 
  \left| 
  \begin{array}{l}
  (a_2,A_2), \ldots, (a_p,A_p) \\
  (b_1,B_1), \ldots, (b_{q-1},B_{q-1}) \\
  \end{array} 
  \right. \right]
\end{align}
This formula is applicable if 
 one of $(a_j, A_j)\; (j=1,\ldots,n)$ is equal to one of the 
 $(b_j, B_j)\; (j=m+1,\ldots,q)$ or 
 one of the $(b_j, B_j)\; (j=1,\ldots,m)$ is equal to one of the $(a_j, A_j)\; (j=n+1,\ldots,p)$, provided that $n\geqslant 1$ and $q \geqslant m$.  

By another round of inverse LT applied on $A_{\gamma}(t)$ given   in Eq.\;\ref{eq:Agammat0} we obtain:
\begin{equation}
D_{\gamma}(\lambda)=\frac{\sin (\gamma \pi   )  }{\pi }  \theta \left(\lambda -{\tau_{\gamma}^{-1}}\right) (\lambda  \tau_{\gamma} -1)^{-\gamma }
\end{equation}
\textcolor{black}{and} \cite{barsoukov2018impedance}
\begin{align}
g_{\gamma}(\tau) =\frac{\sin (\gamma \pi   )  }{\pi  }  \theta \left(\tau^{-1} -{\tau_{\gamma}^{-1}}\right) 
\tau^{-1} (\tau_{\gamma}/\tau -1)^{-\gamma }
\end{align}
i.e.:
\begin{align}
g_{\gamma}(\tau) &=\frac{\sin (\gamma \pi   )  }{\pi   } \tau^{-1}   (\tau_{\gamma}/\tau -1)^{-\gamma }, \;\; \tau< \tau_{\gamma} \nonumber \\
& = 0, \;\; \text{otherwise}
\end{align}
Similarly, we obtain from the $H$-function representation of  
 $A_{\gamma}(t)$ (i.e. Eq.\;\ref{eq:Agammat2}):
 \begin{align}
D_{\gamma}(\lambda) &=  \frac{(\tau_{\gamma} \lambda)^{-1}}{  \Gamma(\gamma)} 
H^{1,1}_{2,2}\left[  (\tau_{\gamma} \lambda)^{-1}   \left|
\begin{array}{l}
(0,1),(0,1)     \\
(\gamma-1,1),(0,1)    \\
\end{array}
\right.\right] \\
&=   \frac{1}{   \Gamma(\gamma)} 
H^{1,1}_{2,2}\left[  (\tau_{\gamma} \lambda)^{-1}   \left|
\begin{array}{l}
(1,1),(1,1)     \\
(\gamma,1),(1,1)    \\
\end{array}
\right.\right]
\end{align}
and
\begin{align}
g_{\gamma}(\tau) =  \frac{1}{ \tau_{\gamma}\, \Gamma(\gamma)} 
%H^{1,1}_{2,2}\left[  \tau /\tau_{\gamma}   \left|
%\begin{array}{c}
%(0,1),(0,1) \hfill    \\
%(\gamma-1,1),(0,1)    \\
%\end{array}
%\right.\right]
H^{1,0}_{1,1}\left[  \tau /\tau_{\gamma}   \left|
\begin{array}{c}
(0,1) \hfill    \\
(\gamma-1,1)   \\
\end{array}
\right.\right]
\label{eq:gDC}
\end{align}

\begin{figure}[!t]
\begin{center}
\includegraphics[width=0.98\columnwidth]{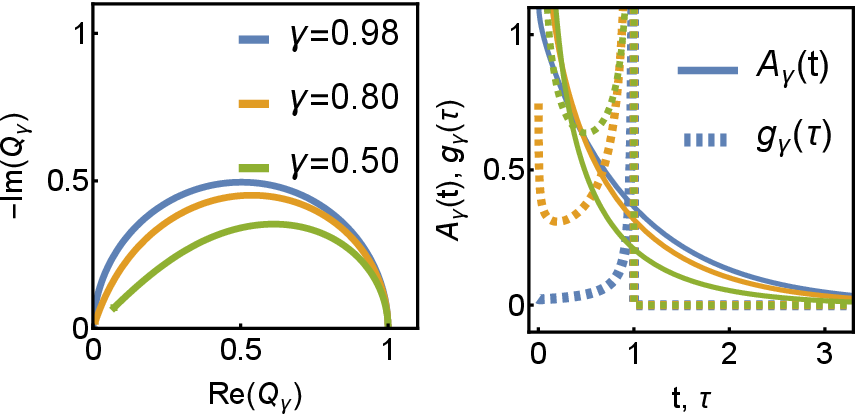}
\caption{Plots of (a) $Q_{\gamma}(s)$ (Eq.\;\ref{eq:QgammaH})  in Nyquist form of real vs. imaginary parts for $\tau_c=1$ and  $\gamma=$ 0.98, 0.80 and 0.50.  
In (b) we plot the functions $A_{\gamma}(t)$ (Eq.\;\ref{eq:Agammat20}) and $g_{\gamma}(\tau)$ (Eq.\;\ref{eq:gDC}) for the same parameters:  $\tau_{\gamma}=1$ and  $\gamma=$ 0.98, 0.80 and 0.50}
\label{fig2}
\end{center}
\end{figure}

For illustration, we show in Fig.\;\ref{fig2}  plots of $Q_{\gamma}(s)$ (Eq.\;\ref{eq:QgammaH}),  $A_{\gamma}(t)$ (Eq.\;\ref{eq:Agammat20}) and $g_{\gamma}(\tau)$ (Eq.\;\ref{eq:gDC}) for $\tau_{\gamma}=1$ and  $\gamma=$ 0.98, 0.80 and 0.50. We see clearly the cutoff of $g_{\gamma}(\tau)$ for values of $\tau>\tau_{\gamma}$. 

\subsection{Cole-Cole model}

 The Cole-Cole model  is given by \cite{cole1941dispersion}:
\begin{equation}
Q_{\alpha}(s) = \frac{1}{1+(s\tau_{\alpha})^{\alpha}} \quad (0<\alpha \leqslant 1)
\label{eq:Qsalpha0}
\end{equation}
in its traditional form, which can be rewritten in terms of the $H$-function  as:
\begin{align}
Q_{\alpha}(s) =  
H^{1,1}_{1,1}\left[ (s\tau_{\alpha})^{\alpha}  \left|
\begin{array}{c}
(0,1)    \\
(0,1)   \\
\end{array}
\right.\right]
\label{eq:Qsalpha}
\end{align} 
This is obtained from\;\cite{mathai2009h}:
\begin{align}
\frac{z^{\beta}}{1+a z^{\alpha}} =  a^{\beta/\alpha}
H^{1,1}_{1,1}\left[ a z^{\alpha}  \left|
\begin{array}{c}
(\beta/\alpha,1)    \\
(\beta/\alpha,1)   \\
\end{array}
\right.\right]
\end{align} 
 The  inverse LT of Eq.\;\ref{eq:Qsalpha0} from the Prabhakar integral \cite{prabhakar1971singular, saxena2004generalized}:
\begin{equation}
  \int \limits_0^{\infty}  t^{\beta-1}  {E}_{\alpha,\beta}^{\gamma} \left( -at^{\alpha}\right) e^{-st} dt =  \frac{s^{-\beta}}{(1+as^{-\alpha})^{\gamma}} %= \frac{s^{\alpha\gamma-\beta}}{(s^{\alpha}+a)^{\gamma}} 
  \label{e5}
\end{equation}
with $\alpha=\beta$, $\gamma=1$ and $a=\tau_{\alpha}^{-\alpha}$ gives the system response function $A_{\alpha}(t)$ as:
\begin{equation}
A_{\alpha}(t) 
=\tau_{\alpha}^{-1}({t}/{\tau_{\alpha}})^{\alpha-1} E_{\alpha,\alpha}^1\left[ -(t/\tau_{\alpha})^{\alpha} \right]
\label{eq48}
\end{equation} 
The function  
\begin{equation}
{E}_{\alpha,\beta}^{\gamma} ( z ) := \sum\limits_{k=0}^{\infty} \frac{(\gamma)_k}{\Gamma(\alpha k + \beta)} \frac{z^k}{k!} \quad (\alpha,\beta, \gamma \in \mathbb{C}, \mathrm{Re}({\alpha})>0)
\label{eqML}
\end{equation}
 with $(\gamma)_k = \gamma(\gamma+1)\ldots(\gamma+k-1) =\Gamma(\gamma+k)/\Gamma(\gamma)$ being the Pochhammer symbol  is the three-parameter Mittag-Leffler  function. 
Eq.\;\ref{eq48} can be expressed in terms of Fox's $H$-function as:
\begin{equation}
A_{\alpha}(t) = \tau_{\alpha}^{-1}({t}/{\tau_{\alpha}})^{\alpha-1} H^{1,1}_{1,2} \left[(t/\tau_{\alpha})^{\alpha}\left|
\begin{array}{l}
{(0,1)} \hfill \\
{(0,1),(1-\alpha,\alpha)} 
\end{array}
\right.
\right]
\label{eq:At1}
\end{equation} 
Using the property\;\ref{H-power}, 
with $\sigma=1-1/\alpha$, Eq.\;\ref{eq:At1} turns to be:
 \begin{equation}
A_{\alpha}(t) = \tau_{\alpha}^{-1} H^{1,1}_{1,2} \left[(t/\tau_{\alpha})^{\alpha} 
\left|
\begin{array}{l}
{(1-1/\alpha,1)} \\
{(1-1/\alpha,1),(0,\alpha)}\\
\end{array} 
\right.
\right]
\label{eq:At12}
\end{equation}
Now we apply the inverse LT formula given in \ref{eq:ILT} on $Q(s)$ given by Eq\;\ref{eq:Qsalpha}. This leads to:
\begin{align}
A_{\alpha}(t)& = \mathcal{L}^{-1} 
\left[ 
H_{1,1}^{1,1}\left[ (s \tau_{\alpha})^{\alpha}  \left|
\begin{array}{c}
(0,1)  \\
(0,1)  \\
\end{array}
\right.\right]; t
\right] \nonumber \\ 
 &= t^{-1} 
H^{1,1}_{2,1}\left[ (t/\tau_{\alpha})^{-\alpha} \left|
\begin{array}{c}
(0,1), (0,\alpha)  \\
(0,1) \hfill  \\
\end{array}
\right.\right]
\label{eq:At36}
\end{align}
With  the use of the property\;\ref{H-inverse},  
$A_{\alpha}(t)$ in Eq\;\ref{eq:At36} becomes:
\begin{equation}
A_{\alpha}(t) = t^{-1} H^{1,1}_{1,2} \left[(t/\tau_{\alpha})^{\alpha}
\left|
\begin{array}{l}
{(1,1)} \\
{(1,1),(1,\alpha)} \\
\end{array} 
\right.
 \right]
 \label{eq:At54}
\end{equation}
and then, with the property\;\ref{H-power} with $\sigma=-1/\alpha$, we  obtain the same result as given in Eq\;\ref{eq:At12}. 

From $A_{\alpha}(t)$ we can now derive $D_{\alpha}(\lambda)$ by another iteration of inverse LT. This gives (after using property\;\ref{H-power}):
\begin{align}
D_{\alpha}(\lambda)& = \mathcal{L}^{-1} 
\left[ 
\tau_{\alpha}^{-1} H^{1,1}_{1,2} \left[(t/\tau_{\alpha})^{\alpha}|^{(1-1/\alpha,1)}_{(1-1/\alpha,1),(0,\alpha)} \right]; \lambda
\right] \nonumber \\ 
 &=  
H^{1,1}_{2,2}\left[ (\tau_{\alpha} \lambda)^{-\alpha} \left|
\begin{array}{c}
(1,1), (1,\alpha)  \\
(1,1), (1,\alpha)   \\
\end{array}
\right.\right]
\label{eq:At39}
\end{align}
With Eq.\;\ref{Eq:g} and then formula\;\ref{H-power}, we obtain for the DFRT $g(\tau)$:
\begin{equation}
g_{\alpha}(\tau) = \tau_{\alpha}^{-1} H^{1,1}_{2,2}\left[ (\tau/\tau_{\alpha})^{\alpha} \left|
\begin{array}{c}
(1-1/\alpha,1), (0,\alpha)  \\
(1-1/\alpha,1), (0,\alpha)   \\
\end{array}
\right.\right]
\label{eq:galphaCC}
\end{equation}

\subsection{Havriliak–Negami model}

Combining both the Cole-Cole and Davidson–Cole models results in the Havriliak–Negami model \cite{havriliak1966complex}:
\begin{align}
Q_{\nu}(s) &= \frac{1}{(1+(s\tau_{{\nu}})^{\alpha})^{\gamma}} \quad (0< \alpha,\gamma \leqslant 1) \\
&= \frac{1}{\Gamma(\gamma)} 
H^{1,1}_{1,1}\left[ (s\tau_{\nu})^{\alpha}  \left|
\begin{array}{c}
(1-\gamma,1)    \\
(0,1) \hfill   \\
\end{array}
\right.\right]
\label{eq:QsalphaHN}
\end{align}
It is easy to verify that for the special case of $\gamma=1$, Eq.\;\ref{eq:QsalphaHN} reduces to the Cole-Cole relation (Eq.\;\ref{eq:Qsalpha}), and  in the case that $\alpha=1$, the Davidson-Cole relation is obtained (Eq.\;\ref{eq:QgammaH}). 
By operating a first LT inversion on $Q_{\nu}(s)$ (using  Eq.\;\ref{e5} and Eq.\;\ref{eq:ILT}) we obtain:
\begin{align}
A_{\nu}(t) 
&=\tau_{\nu}^{-1}({t}/{\tau_{\nu}})^{\alpha\gamma-1} E_{\alpha,\alpha\gamma}^{\gamma}\left[ -(t/\tau_{\nu})^{\alpha} \right] \\
&= \frac{1}{t\, \Gamma(\gamma)} 
H^{1,1}_{1,2}\left[ (t/\tau_{\nu})^{\alpha}  \left|
\begin{array}{c}
(1,1)   \hfill \\
(\gamma,1), (1,\alpha)    \\
\end{array}
\right.\right]
\label{eq41}
\end{align}
and then again on $A_{\nu}(t)$ with the use of Eq.\;\ref{Eq:g} leads to:
\begin{equation}
g_{\nu}(\tau) = \frac{1}{\tau_{\nu} \Gamma(\gamma)} H^{1,1}_{2,2}\left[ (\tau/\tau_{\nu})^{\alpha} \left|
\begin{array}{c}
(1-1/\alpha,1), (0,\alpha)  \\
(\gamma-1/\alpha,1), (0,\alpha)   \\
\end{array}
\right.\right]
\label{eq:gHN}
\end{equation}
We verify that the response function in 
Eq.\;\ref{eq41}   reduces to that given by Eq.\;\ref{eq:At54} for $\gamma=1$ (Cole-Cole model), and to Eq.\;\ref{eq:Agammat20} for $\alpha=1$ (Davidson-Cole model). The DFRT in 
Eq.\;\ref{eq:gHN} reduces to Eq.\;\ref{eq:galphaCC} for $\gamma=1$, and to Eq.\;\ref{eq:gDC} for $\alpha=1$.

\begin{figure}[!t]
\begin{center}
\includegraphics[width=0.98\columnwidth]{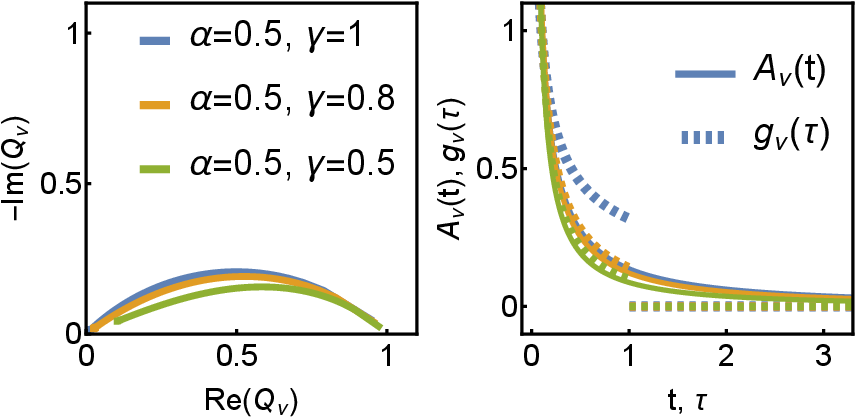}
\caption{
Plots of (a) $Q_{\nu}(s)$ (Eq.\;\ref{eq:QsalphaHN})  in Nyquist form of real vs. imaginary parts for $\tau_{\nu}=1$, $\alpha=0.5$ and  $\gamma=$ 1, 0.80 and 0.50.  
In (b) we plot the functions 
$A_{\nu}(t)$ (Eq.\;\ref{eq41})
 and $g_{\nu}(\tau)$ (Eq.\;\ref{eq:gHN})
  for the same parameters $\tau_{\nu}$, $\alpha$ and  $\gamma$
}
\label{fig3}
\end{center}
\end{figure}

Plots of $Q_{\nu}(s)$ (Eq.\;\ref{eq:QsalphaHN}), $A_{\nu}(t)$  (Eq.\;\ref{eq41}), and $g_{\tau}(s)$ (Eq.\;\ref{eq:gHN}) for $\tau_{\nu}=1$ and a combination of values of $\alpha$ and $\gamma$  (Cole-Cole model for $\gamma=1$) are given in Fig.\;\ref{fig3}.

\section{Conclusion}

With the assumption that a system can be represented by an infinite number of $RC$ branches, the distribution function of the $RC$ time constants  is related to the impedance model via two iterated Laplace transforms. 
We showed in this paper a general procedure for how to deconvolve the DFRT starting from any arbitrary  impedance model represented in terms of the Fox $H$-function.

\section*{References}

%\bibliography{bib}

\begin{thebibliography}{45}%
\makeatletter
\providecommand \@ifxundefined [1]{%
 \@ifx{#1\undefined}
}%
\providecommand \@ifnum [1]{%
 \ifnum #1\expandafter \@firstoftwo
 \else \expandafter \@secondoftwo
 \fi
}%
\providecommand \@ifx [1]{%
 \ifx #1\expandafter \@firstoftwo
 \else \expandafter \@secondoftwo
 \fi
}%
\providecommand \natexlab [1]{#1}%
\providecommand \enquote  [1]{``#1''}%
\providecommand \bibnamefont  [1]{#1}%
\providecommand \bibfnamefont [1]{#1}%
\providecommand \citenamefont [1]{#1}%
\providecommand \href@noop [0]{\@secondoftwo}%
\providecommand \href [0]{\begingroup \@sanitize@url \@href}%
\providecommand \@href[1]{\@@startlink{#1}\@@href}%
\providecommand \@@href[1]{\endgroup#1\@@endlink}%
\providecommand \@sanitize@url [0]{\catcode `\\12\catcode `\$12\catcode
  `\&12\catcode `\#12\catcode `\^12\catcode `\_12\catcode `\%12\relax}%
\providecommand \@@startlink[1]{}%
\providecommand \@@endlink[0]{}%
\providecommand \url  [0]{\begingroup\@sanitize@url \@url }%
\providecommand \@url [1]{\endgroup\@href {#1}{\urlprefix }}%
\providecommand \urlprefix  [0]{URL }%
\providecommand \Eprint [0]{\href }%
\providecommand \doibase [0]{http://dx.doi.org/}%
\providecommand \selectlanguage [0]{\@gobble}%
\providecommand \bibinfo  [0]{\@secondoftwo}%
\providecommand \bibfield  [0]{\@secondoftwo}%
\providecommand \translation [1]{[#1]}%
\providecommand \BibitemOpen [0]{}%
\providecommand \bibitemStop [0]{}%
\providecommand \bibitemNoStop [0]{.\EOS\space}%
\providecommand \EOS [0]{\spacefactor3000\relax}%
\providecommand \BibitemShut  [1]{\csname bibitem#1\endcsname}%
\let\auto@bib@innerbib\@empty
%</preamble>
\bibitem [{\citenamefont {Barsoukov}\ and\ \citenamefont
  {Macdonald}(2018)}]{barsoukov2018impedance}%
  \BibitemOpen
  \bibfield  {author} {\bibinfo {author} {\bibfnamefont {E.}~\bibnamefont
  {Barsoukov}}\ and\ \bibinfo {author} {\bibfnamefont {J.~R.}\ \bibnamefont
  {Macdonald}},\ }\href@noop {} {\emph {\bibinfo {title} {Impedance
  spectroscopy: theory, experiment, and applications}}}\ (\bibinfo  {publisher}
  {John Wiley \& Sons},\ \bibinfo {year} {2018})\BibitemShut {NoStop}%
\bibitem [{\citenamefont {Lasia}(2014)}]{lasia2014electrochemical}%
  \BibitemOpen
  \bibfield  {author} {\bibinfo {author} {\bibfnamefont {A.}~\bibnamefont
  {Lasia}},\ }\href@noop {} {\emph {\bibinfo {title} {Electrochemical impedance
  spectroscopy and its applications}}}\ (\bibinfo  {publisher} {Springer},\
  \bibinfo {year} {2014})\BibitemShut {NoStop}%
\bibitem [{\citenamefont {Single}, \citenamefont {Horstmann},\ and\
  \citenamefont {Latz}(2019)}]{single2019theory}%
  \BibitemOpen
  \bibfield  {author} {\bibinfo {author} {\bibfnamefont {F.}~\bibnamefont
  {Single}}, \bibinfo {author} {\bibfnamefont {B.}~\bibnamefont {Horstmann}}, \
  and\ \bibinfo {author} {\bibfnamefont {A.}~\bibnamefont {Latz}},\ }\bibfield
  {title} {\enquote {\bibinfo {title} {Theory of impedance spectroscopy for
  lithium batteries},}\ }\href@noop {} {\bibfield  {journal} {\bibinfo
  {journal} {J. Phys. Chem. C}\ }\textbf {\bibinfo {volume} {123}},\ \bibinfo
  {pages} {27327--27343} (\bibinfo {year} {2019})}\BibitemShut {NoStop}%
\bibitem [{\citenamefont {Bredar}\ \emph {et~al.}(2020)\citenamefont {Bredar},
  \citenamefont {Chown}, \citenamefont {Burton},\ and\ \citenamefont
  {Farnum}}]{bredar2020electrochemical}%
  \BibitemOpen
  \bibfield  {author} {\bibinfo {author} {\bibfnamefont {A.~R.}\ \bibnamefont
  {Bredar}}, \bibinfo {author} {\bibfnamefont {A.~L.}\ \bibnamefont {Chown}},
  \bibinfo {author} {\bibfnamefont {A.~R.}\ \bibnamefont {Burton}}, \ and\
  \bibinfo {author} {\bibfnamefont {B.~H.}\ \bibnamefont {Farnum}},\ }\bibfield
   {title} {\enquote {\bibinfo {title} {Electrochemical impedance spectroscopy
  of metal oxide electrodes for energy applications},}\ }\href@noop {}
  {\bibfield  {journal} {\bibinfo  {journal} {ACS Appl. Energy Mater.}\
  }\textbf {\bibinfo {volume} {3}},\ \bibinfo {pages} {66--98} (\bibinfo {year}
  {2020})}\BibitemShut {NoStop}%
\bibitem [{\citenamefont {Vadhva}\ \emph {et~al.}(2021)\citenamefont {Vadhva},
  \citenamefont {Hu}, \citenamefont {Johnson}, \citenamefont {Stocker},
  \citenamefont {Braglia}, \citenamefont {Brett},\ and\ \citenamefont
  {Rettie}}]{vadhva2021electrochemical}%
  \BibitemOpen
  \bibfield  {author} {\bibinfo {author} {\bibfnamefont {P.}~\bibnamefont
  {Vadhva}}, \bibinfo {author} {\bibfnamefont {J.}~\bibnamefont {Hu}}, \bibinfo
  {author} {\bibfnamefont {M.~J.}\ \bibnamefont {Johnson}}, \bibinfo {author}
  {\bibfnamefont {R.}~\bibnamefont {Stocker}}, \bibinfo {author} {\bibfnamefont
  {M.}~\bibnamefont {Braglia}}, \bibinfo {author} {\bibfnamefont {D.~J.}\
  \bibnamefont {Brett}}, \ and\ \bibinfo {author} {\bibfnamefont {A.~J.}\
  \bibnamefont {Rettie}},\ }\bibfield  {title} {\enquote {\bibinfo {title}
  {Electrochemical impedance spectroscopy for all-solid-state batteries:
  Theory, methods and future outlook},}\ }\href@noop {} {\bibfield  {journal}
  {\bibinfo  {journal} {ChemElectroChem}\ }\textbf {\bibinfo {volume} {8}},\
  \bibinfo {pages} {1930--1947} (\bibinfo {year} {2021})}\BibitemShut {NoStop}%
\bibitem [{\citenamefont {Allagui}\ \emph {et~al.}(2016)\citenamefont
  {Allagui}, \citenamefont {Elwakil}, \citenamefont {Maundy},\ and\
  \citenamefont {Freeborn}}]{eis}%
  \BibitemOpen
  \bibfield  {author} {\bibinfo {author} {\bibfnamefont {A.}~\bibnamefont
  {Allagui}}, \bibinfo {author} {\bibfnamefont {A.~S.}\ \bibnamefont
  {Elwakil}}, \bibinfo {author} {\bibfnamefont {B.~J.}\ \bibnamefont {Maundy}},
  \ and\ \bibinfo {author} {\bibfnamefont {T.~J.}\ \bibnamefont {Freeborn}},\
  }\bibfield  {title} {\enquote {\bibinfo {title} {Spectral capacitance of
  series and parallel combinations of supercapacitors},}\ }\href@noop {}
  {\bibfield  {journal} {\bibinfo  {journal} {ChemElectroChem}\ }\textbf
  {\bibinfo {volume} {3}},\ \bibinfo {pages} {1429--1436} (\bibinfo {year}
  {2016})}\BibitemShut {NoStop}%
\bibitem [{\citenamefont {He}\ and\ \citenamefont
  {Mansfeld}(2009)}]{he2009exploring}%
  \BibitemOpen
  \bibfield  {author} {\bibinfo {author} {\bibfnamefont {Z.}~\bibnamefont
  {He}}\ and\ \bibinfo {author} {\bibfnamefont {F.}~\bibnamefont {Mansfeld}},\
  }\bibfield  {title} {\enquote {\bibinfo {title} {Exploring the use of
  electrochemical impedance spectroscopy (eis) in microbial fuel cell
  studies},}\ }\href@noop {} {\bibfield  {journal} {\bibinfo  {journal} {Energy
  Environ. Sci.}\ }\textbf {\bibinfo {volume} {2}},\ \bibinfo {pages}
  {215--219} (\bibinfo {year} {2009})}\BibitemShut {NoStop}%
\bibitem [{\citenamefont {Guerrero}, \citenamefont {Bisquert},\ and\
  \citenamefont {Garcia-Belmonte}(2021)}]{guerrero2021impedance}%
  \BibitemOpen
  \bibfield  {author} {\bibinfo {author} {\bibfnamefont {A.}~\bibnamefont
  {Guerrero}}, \bibinfo {author} {\bibfnamefont {J.}~\bibnamefont {Bisquert}},
  \ and\ \bibinfo {author} {\bibfnamefont {G.}~\bibnamefont
  {Garcia-Belmonte}},\ }\bibfield  {title} {\enquote {\bibinfo {title}
  {Impedance spectroscopy of metal halide perovskite solar cells from the
  perspective of equivalent circuits},}\ }\href@noop {} {\bibfield  {journal}
  {\bibinfo  {journal} {Chem. Rev.}\ }\textbf {\bibinfo {volume} {121}},\
  \bibinfo {pages} {14430--14484} (\bibinfo {year} {2021})}\BibitemShut
  {NoStop}%
\bibitem [{\citenamefont {Pejcic}\ and\ \citenamefont
  {De~Marco}(2006)}]{pejcic2006impedance}%
  \BibitemOpen
  \bibfield  {author} {\bibinfo {author} {\bibfnamefont {B.}~\bibnamefont
  {Pejcic}}\ and\ \bibinfo {author} {\bibfnamefont {R.}~\bibnamefont
  {De~Marco}},\ }\bibfield  {title} {\enquote {\bibinfo {title} {Impedance
  spectroscopy: Over 35 years of electrochemical sensor optimization},}\
  }\href@noop {} {\bibfield  {journal} {\bibinfo  {journal} {Electrochim.
  Acta}\ }\textbf {\bibinfo {volume} {51}},\ \bibinfo {pages} {6217--6229}
  (\bibinfo {year} {2006})}\BibitemShut {NoStop}%
\bibitem [{\citenamefont {Mansfeld}(1990)}]{mansfeld1990electrochemical}%
  \BibitemOpen
  \bibfield  {author} {\bibinfo {author} {\bibfnamefont {F.}~\bibnamefont
  {Mansfeld}},\ }\bibfield  {title} {\enquote {\bibinfo {title}
  {Electrochemical impedance spectroscopy (eis) as a new tool for investigating
  methods of corrosion protection},}\ }\href@noop {} {\bibfield  {journal}
  {\bibinfo  {journal} {Electrochim. Acta}\ }\textbf {\bibinfo {volume} {35}},\
  \bibinfo {pages} {1533--1544} (\bibinfo {year} {1990})}\BibitemShut {NoStop}%
\bibitem [{\citenamefont {Cesewski}\ and\ \citenamefont
  {Johnson}(2020)}]{CESEWSKI2020112214}%
  \BibitemOpen
  \bibfield  {author} {\bibinfo {author} {\bibfnamefont {E.}~\bibnamefont
  {Cesewski}}\ and\ \bibinfo {author} {\bibfnamefont {B.~N.}\ \bibnamefont
  {Johnson}},\ }\bibfield  {title} {\enquote {\bibinfo {title} {Electrochemical
  biosensors for pathogen detection},}\ }\href@noop {} {\bibfield  {journal}
  {\bibinfo  {journal} {Biosens. Bioelectron.}\ }\textbf {\bibinfo {volume}
  {159}},\ \bibinfo {pages} {112214} (\bibinfo {year} {2020})}\BibitemShut
  {NoStop}%
\bibitem [{\citenamefont {Lisdat}\ and\ \citenamefont
  {Sch{\"a}fer}(2008)}]{lisdat2008use}%
  \BibitemOpen
  \bibfield  {author} {\bibinfo {author} {\bibfnamefont {F.}~\bibnamefont
  {Lisdat}}\ and\ \bibinfo {author} {\bibfnamefont {D.}~\bibnamefont
  {Sch{\"a}fer}},\ }\bibfield  {title} {\enquote {\bibinfo {title} {The use of
  electrochemical impedance spectroscopy for biosensing},}\ }\href@noop {}
  {\bibfield  {journal} {\bibinfo  {journal} {Anal. Bioanal.Chem.}\ }\textbf
  {\bibinfo {volume} {391}},\ \bibinfo {pages} {1555--1567} (\bibinfo {year}
  {2008})}\BibitemShut {NoStop}%
\bibitem [{\citenamefont {Effendy}, \citenamefont {Song},\ and\ \citenamefont
  {Bazant}(2020)}]{effendy2020analysis}%
  \BibitemOpen
  \bibfield  {author} {\bibinfo {author} {\bibfnamefont {S.}~\bibnamefont
  {Effendy}}, \bibinfo {author} {\bibfnamefont {J.}~\bibnamefont {Song}}, \
  and\ \bibinfo {author} {\bibfnamefont {M.~Z.}\ \bibnamefont {Bazant}},\
  }\bibfield  {title} {\enquote {\bibinfo {title} {Analysis, design, and
  generalization of electrochemical impedance spectroscopy (eis) inversion
  algorithms},}\ }\href@noop {} {\bibfield  {journal} {\bibinfo  {journal} {J.
  Electrochem. Soc.}\ }\textbf {\bibinfo {volume} {167}},\ \bibinfo {pages}
  {106508} (\bibinfo {year} {2020})}\BibitemShut {NoStop}%
\bibitem [{\citenamefont {Ciucci}\ and\ \citenamefont
  {Chen}(2015)}]{ciucci2015analysis}%
  \BibitemOpen
  \bibfield  {author} {\bibinfo {author} {\bibfnamefont {F.}~\bibnamefont
  {Ciucci}}\ and\ \bibinfo {author} {\bibfnamefont {C.}~\bibnamefont {Chen}},\
  }\bibfield  {title} {\enquote {\bibinfo {title} {Analysis of electrochemical
  impedance spectroscopy data using the distribution of relaxation times: A
  bayesian and hierarchical bayesian approach},}\ }\href@noop {} {\bibfield
  {journal} {\bibinfo  {journal} {Electrochim. Acta}\ }\textbf {\bibinfo
  {volume} {167}},\ \bibinfo {pages} {439--454} (\bibinfo {year}
  {2015})}\BibitemShut {NoStop}%
\bibitem [{\citenamefont {Allagui}\ and\ \citenamefont
  {Fouda}(2021)}]{allagui2021inverse}%
  \BibitemOpen
  \bibfield  {author} {\bibinfo {author} {\bibfnamefont {A.}~\bibnamefont
  {Allagui}}\ and\ \bibinfo {author} {\bibfnamefont {M.~E.}\ \bibnamefont
  {Fouda}},\ }\bibfield  {title} {\enquote {\bibinfo {title} {Inverse problem
  of reconstructing the capacitance of electric double-layer capacitors},}\
  }\href@noop {} {\bibfield  {journal} {\bibinfo  {journal} {Electrochim.
  Acta}\ ,\ \bibinfo {pages} {138848}} (\bibinfo {year} {2021})}\BibitemShut
  {NoStop}%
\bibitem [{\citenamefont {Allagui}, \citenamefont {Elwakil},\ and\
  \citenamefont {Fouda}(2021)}]{ieeeted}%
  \BibitemOpen
  \bibfield  {author} {\bibinfo {author} {\bibfnamefont {A.}~\bibnamefont
  {Allagui}}, \bibinfo {author} {\bibfnamefont {A.~S.}\ \bibnamefont
  {Elwakil}}, \ and\ \bibinfo {author} {\bibfnamefont {M.~E.}\ \bibnamefont
  {Fouda}},\ }\bibfield  {title} {\enquote {\bibinfo {title} {Revisiting the
  time-domain and frequency-domain definitions of capacitance},}\ }\href@noop
  {} {\bibfield  {journal} {\bibinfo  {journal} {IEEE Trans. Electron Devices}\
  }\textbf {\bibinfo {volume} {68}} (\bibinfo {year} {2021})}\BibitemShut
  {NoStop}%
\bibitem [{\citenamefont {Florsch}, \citenamefont {Revil},\ and\ \citenamefont
  {Camerlynck}(2014)}]{florsch2014inversion}%
  \BibitemOpen
  \bibfield  {author} {\bibinfo {author} {\bibfnamefont {N.}~\bibnamefont
  {Florsch}}, \bibinfo {author} {\bibfnamefont {A.}~\bibnamefont {Revil}}, \
  and\ \bibinfo {author} {\bibfnamefont {C.}~\bibnamefont {Camerlynck}},\
  }\bibfield  {title} {\enquote {\bibinfo {title} {Inversion of generalized
  relaxation time distributions with optimized damping parameter},}\
  }\href@noop {} {\bibfield  {journal} {\bibinfo  {journal} {J. Appl.
  Geophys.}\ }\textbf {\bibinfo {volume} {109}},\ \bibinfo {pages} {119--132}
  (\bibinfo {year} {2014})}\BibitemShut {NoStop}%
\bibitem [{\citenamefont {Schichlein}\ \emph {et~al.}(2002)\citenamefont
  {Schichlein}, \citenamefont {M{\"u}ller}, \citenamefont {Voigts},
  \citenamefont {Kr{\"u}gel},\ and\ \citenamefont
  {Ivers-Tiff{\'e}e}}]{schichlein2002deconvolution}%
  \BibitemOpen
  \bibfield  {author} {\bibinfo {author} {\bibfnamefont {H.}~\bibnamefont
  {Schichlein}}, \bibinfo {author} {\bibfnamefont {A.~C.}\ \bibnamefont
  {M{\"u}ller}}, \bibinfo {author} {\bibfnamefont {M.}~\bibnamefont {Voigts}},
  \bibinfo {author} {\bibfnamefont {A.}~\bibnamefont {Kr{\"u}gel}}, \ and\
  \bibinfo {author} {\bibfnamefont {E.}~\bibnamefont {Ivers-Tiff{\'e}e}},\
  }\bibfield  {title} {\enquote {\bibinfo {title} {Deconvolution of
  electrochemical impedance spectra for the identification of electrode
  reaction mechanisms in solid oxide fuel cells},}\ }\href@noop {} {\bibfield
  {journal} {\bibinfo  {journal} {J. Appl. Electrochem.}\ }\textbf {\bibinfo
  {volume} {32}},\ \bibinfo {pages} {875--882} (\bibinfo {year}
  {2002})}\BibitemShut {NoStop}%
\bibitem [{\citenamefont {Boukamp}(2015)}]{boukamp2015fourier}%
  \BibitemOpen
  \bibfield  {author} {\bibinfo {author} {\bibfnamefont {B.~A.}\ \bibnamefont
  {Boukamp}},\ }\bibfield  {title} {\enquote {\bibinfo {title} {Fourier
  transform distribution function of relaxation times; application and
  limitations},}\ }\href@noop {} {\bibfield  {journal} {\bibinfo  {journal}
  {Electrochimica acta}\ }\textbf {\bibinfo {volume} {154}},\ \bibinfo {pages}
  {35--46} (\bibinfo {year} {2015})}\BibitemShut {NoStop}%
\bibitem [{\citenamefont {Gavrilyuk}, \citenamefont {Osinkin},\ and\
  \citenamefont {Bronin}(2017)}]{gavrilyuk2017use}%
  \BibitemOpen
  \bibfield  {author} {\bibinfo {author} {\bibfnamefont {A.}~\bibnamefont
  {Gavrilyuk}}, \bibinfo {author} {\bibfnamefont {D.}~\bibnamefont {Osinkin}},
  \ and\ \bibinfo {author} {\bibfnamefont {D.}~\bibnamefont {Bronin}},\
  }\bibfield  {title} {\enquote {\bibinfo {title} {The use of tikhonov
  regularization method for calculating the distribution function of relaxation
  times in impedance spectroscopy},}\ }\href@noop {} {\bibfield  {journal}
  {\bibinfo  {journal} {Russ. J. Electrochem.}\ }\textbf {\bibinfo {volume}
  {53}},\ \bibinfo {pages} {575--588} (\bibinfo {year} {2017})}\BibitemShut
  {NoStop}%
\bibitem [{\citenamefont {Paul}\ \emph {et~al.}(2021)\citenamefont {Paul},
  \citenamefont {Chi}, \citenamefont {Wu},\ and\ \citenamefont
  {Wu}}]{paul2021computation}%
  \BibitemOpen
  \bibfield  {author} {\bibinfo {author} {\bibfnamefont {T.}~\bibnamefont
  {Paul}}, \bibinfo {author} {\bibfnamefont {P.}~\bibnamefont {Chi}}, \bibinfo
  {author} {\bibfnamefont {P.~M.}\ \bibnamefont {Wu}}, \ and\ \bibinfo {author}
  {\bibfnamefont {M.}~\bibnamefont {Wu}},\ }\bibfield  {title} {\enquote
  {\bibinfo {title} {Computation of distribution of relaxation times by
  tikhonov regularization for li ion batteries: usage of l-curve method},}\
  }\href@noop {} {\bibfield  {journal} {\bibinfo  {journal} {Sci. Rep.}\
  }\textbf {\bibinfo {volume} {11}},\ \bibinfo {pages} {12624} (\bibinfo {year}
  {2021})}\BibitemShut {NoStop}%
\bibitem [{\citenamefont {Shanbhag}(2020)}]{shanbhag2020relaxation}%
  \BibitemOpen
  \bibfield  {author} {\bibinfo {author} {\bibfnamefont {S.}~\bibnamefont
  {Shanbhag}},\ }\bibfield  {title} {\enquote {\bibinfo {title} {Relaxation
  spectra using nonlinear tikhonov regularization with a bayesian criterion},}\
  }\href@noop {} {\bibfield  {journal} {\bibinfo  {journal} {Rheol. Acta}\
  }\textbf {\bibinfo {volume} {59}},\ \bibinfo {pages} {509--520} (\bibinfo
  {year} {2020})}\BibitemShut {NoStop}%
\bibitem [{\citenamefont {H{\"o}rlin}(1998)}]{horlin1998deconvolution}%
  \BibitemOpen
  \bibfield  {author} {\bibinfo {author} {\bibfnamefont {T.}~\bibnamefont
  {H{\"o}rlin}},\ }\bibfield  {title} {\enquote {\bibinfo {title}
  {Deconvolution and maximum entropy in impedance spectroscopy of noninductive
  systems},}\ }\href@noop {} {\bibfield  {journal} {\bibinfo  {journal} {Solid
  State Ionics}\ }\textbf {\bibinfo {volume} {107}},\ \bibinfo {pages}
  {241--253} (\bibinfo {year} {1998})}\BibitemShut {NoStop}%
\bibitem [{\citenamefont {H{\"o}rlin}(1993)}]{horlin1993maximum}%
  \BibitemOpen
  \bibfield  {author} {\bibinfo {author} {\bibfnamefont {T.}~\bibnamefont
  {H{\"o}rlin}},\ }\bibfield  {title} {\enquote {\bibinfo {title} {Maximum
  entropy in impedance spectroscopy of non-inductive systems},}\ }\href@noop {}
  {\bibfield  {journal} {\bibinfo  {journal} {Solid State Ionics}\ }\textbf
  {\bibinfo {volume} {67}},\ \bibinfo {pages} {85--96} (\bibinfo {year}
  {1993})}\BibitemShut {NoStop}%
\bibitem [{\citenamefont {Tesler}\ \emph {et~al.}(2010)\citenamefont {Tesler},
  \citenamefont {Lewin}, \citenamefont {Baltianski},\ and\ \citenamefont
  {Tsur}}]{tesler2010analyzing}%
  \BibitemOpen
  \bibfield  {author} {\bibinfo {author} {\bibfnamefont {A.}~\bibnamefont
  {Tesler}}, \bibinfo {author} {\bibfnamefont {D.}~\bibnamefont {Lewin}},
  \bibinfo {author} {\bibfnamefont {S.}~\bibnamefont {Baltianski}}, \ and\
  \bibinfo {author} {\bibfnamefont {Y.}~\bibnamefont {Tsur}},\ }\bibfield
  {title} {\enquote {\bibinfo {title} {Analyzing results of impedance
  spectroscopy using novel evolutionary programming techniques},}\ }\href@noop
  {} {\bibfield  {journal} {\bibinfo  {journal} {J. Electroceram.}\ }\textbf
  {\bibinfo {volume} {24}},\ \bibinfo {pages} {245--260} (\bibinfo {year}
  {2010})}\BibitemShut {NoStop}%
\bibitem [{\citenamefont {Boukamp}(2020)}]{boukamp2020distribution}%
  \BibitemOpen
  \bibfield  {author} {\bibinfo {author} {\bibfnamefont {B.~A.}\ \bibnamefont
  {Boukamp}},\ }\bibfield  {title} {\enquote {\bibinfo {title} {Distribution
  (function) of relaxation times, successor to complex nonlinear least squares
  analysis of electrochemical impedance spectroscopy?}}\ }\href@noop {}
  {\bibfield  {journal} {\bibinfo  {journal} {J. Phys.: Energy}\ }\textbf
  {\bibinfo {volume} {2}},\ \bibinfo {pages} {042001} (\bibinfo {year}
  {2020})}\BibitemShut {NoStop}%
\bibitem [{\citenamefont {Liu}\ and\ \citenamefont
  {Ciucci}(2020)}]{liu2020deep}%
  \BibitemOpen
  \bibfield  {author} {\bibinfo {author} {\bibfnamefont {J.}~\bibnamefont
  {Liu}}\ and\ \bibinfo {author} {\bibfnamefont {F.}~\bibnamefont {Ciucci}},\
  }\bibfield  {title} {\enquote {\bibinfo {title} {The deep-prior distribution
  of relaxation times},}\ }\href@noop {} {\bibfield  {journal} {\bibinfo
  {journal} {J. Electrochem. Soc.}\ }\textbf {\bibinfo {volume} {167}},\
  \bibinfo {pages} {026506} (\bibinfo {year} {2020})}\BibitemShut {NoStop}%
\bibitem [{\citenamefont {Macdonald}\ and\ \citenamefont
  {Brachman}(1956)}]{macdonald1956linear}%
  \BibitemOpen
  \bibfield  {author} {\bibinfo {author} {\bibfnamefont {J.~R.}\ \bibnamefont
  {Macdonald}}\ and\ \bibinfo {author} {\bibfnamefont {M.~K.}\ \bibnamefont
  {Brachman}},\ }\bibfield  {title} {\enquote {\bibinfo {title} {Linear-system
  integral transform relations},}\ }\href@noop {} {\bibfield  {journal}
  {\bibinfo  {journal} {Reviews of modern physics}\ }\textbf {\bibinfo {volume}
  {28}},\ \bibinfo {pages} {393} (\bibinfo {year} {1956})}\BibitemShut
  {NoStop}%
\bibitem [{\citenamefont {Bateman}(1954)}]{bateman1954tables}%
  \BibitemOpen
  \bibfield  {author} {\bibinfo {author} {\bibfnamefont {H.}~\bibnamefont
  {Bateman}},\ }\href@noop {} {\emph {\bibinfo {title} {Tables of integral
  transforms}}},\ Vol.~\bibinfo {volume} {2}\ (\bibinfo  {publisher}
  {McGraw-Hill book company},\ \bibinfo {year} {1954})\BibitemShut {NoStop}%
\bibitem [{\citenamefont {Fox}(1961)}]{fox1961g}%
  \BibitemOpen
  \bibfield  {author} {\bibinfo {author} {\bibfnamefont {C.}~\bibnamefont
  {Fox}},\ }\bibfield  {title} {\enquote {\bibinfo {title} {The g and h
  functions as symmetrical fourier kernels},}\ }\href@noop {} {\bibfield
  {journal} {\bibinfo  {journal} {Trans. Am. Math. Soc.}\ }\textbf {\bibinfo
  {volume} {98}},\ \bibinfo {pages} {395--429} (\bibinfo {year}
  {1961})}\BibitemShut {NoStop}%
\bibitem [{\citenamefont {Mathai}, \citenamefont {Saxena},\ and\ \citenamefont
  {Haubold}(2009)}]{mathai2009h}%
  \BibitemOpen
  \bibfield  {author} {\bibinfo {author} {\bibfnamefont {A.~M.}\ \bibnamefont
  {Mathai}}, \bibinfo {author} {\bibfnamefont {R.~K.}\ \bibnamefont {Saxena}},
  \ and\ \bibinfo {author} {\bibfnamefont {H.~J.}\ \bibnamefont {Haubold}},\
  }\href@noop {} {\emph {\bibinfo {title} {The H-function: theory and
  applications}}}\ (\bibinfo  {publisher} {Springer Science \& Business
  Media},\ \bibinfo {year} {2009})\BibitemShut {NoStop}%
\bibitem [{\citenamefont {Mathai}\ \emph {et~al.}(1978)\citenamefont {Mathai},
  \citenamefont {Saxena}, \citenamefont {Saxena} \emph {et~al.}}]{mathai1978h}%
  \BibitemOpen
  \bibfield  {author} {\bibinfo {author} {\bibfnamefont {A.~M.}\ \bibnamefont
  {Mathai}}, \bibinfo {author} {\bibfnamefont {R.~K.}\ \bibnamefont {Saxena}},
  \bibinfo {author} {\bibfnamefont {R.~K.}\ \bibnamefont {Saxena}},  \emph
  {et~al.},\ }\href@noop {} {\emph {\bibinfo {title} {The H-function with
  applications in statistics and other disciplines}}}\ (\bibinfo  {publisher}
  {John Wiley \& Sons},\ \bibinfo {year} {1978})\BibitemShut {NoStop}%
\bibitem [{\citenamefont {Kilbas}(2004)}]{kilbas2004h}%
  \BibitemOpen
  \bibfield  {author} {\bibinfo {author} {\bibfnamefont {A.~A.}\ \bibnamefont
  {Kilbas}},\ }\href@noop {} {\emph {\bibinfo {title} {H-transforms: Theory and
  Applications}}}\ (\bibinfo  {publisher} {CRC press},\ \bibinfo {year}
  {2004})\BibitemShut {NoStop}%
\bibitem [{\citenamefont {Hilfer}(2002)}]{hilfer}%
  \BibitemOpen
  \bibfield  {author} {\bibinfo {author} {\bibfnamefont {R.}~\bibnamefont
  {Hilfer}},\ }\bibfield  {title} {\enquote {\bibinfo {title} {$h$-function
  representations for stretched exponential relaxation and non-debye
  susceptibilities in glassy systems},}\ }\href@noop {} {\bibfield  {journal}
  {\bibinfo  {journal} {Phys. Rev. E}\ }\textbf {\bibinfo {volume} {65}},\
  \bibinfo {pages} {061510} (\bibinfo {year} {2002})}\BibitemShut {NoStop}%
\bibitem [{\citenamefont {Gl{\"o}ckle}\ and\ \citenamefont
  {Nonnenmacher}(1993)}]{glockle1993fox}%
  \BibitemOpen
  \bibfield  {author} {\bibinfo {author} {\bibfnamefont {W.~G.}\ \bibnamefont
  {Gl{\"o}ckle}}\ and\ \bibinfo {author} {\bibfnamefont {T.~F.}\ \bibnamefont
  {Nonnenmacher}},\ }\bibfield  {title} {\enquote {\bibinfo {title} {Fox
  function representation of non-debye relaxation processes},}\ }\href@noop {}
  {\bibfield  {journal} {\bibinfo  {journal} {J. Stat. Phys.}\ }\textbf
  {\bibinfo {volume} {71}},\ \bibinfo {pages} {741--757} (\bibinfo {year}
  {1993})}\BibitemShut {NoStop}%
\bibitem [{\citenamefont {Lasia}(2022)}]{lasia2022origin}%
  \BibitemOpen
  \bibfield  {author} {\bibinfo {author} {\bibfnamefont {A.}~\bibnamefont
  {Lasia}},\ }\bibfield  {title} {\enquote {\bibinfo {title} {The origin of the
  constant phase element},}\ }\href@noop {} {\bibfield  {journal} {\bibinfo
  {journal} {J. Phys. Chem. Lett.}\ }\textbf {\bibinfo {volume} {13}},\
  \bibinfo {pages} {580--589} (\bibinfo {year} {2022})}\BibitemShut {NoStop}%
\bibitem [{\citenamefont {Gateman}\ \emph {et~al.}(2022)\citenamefont
  {Gateman}, \citenamefont {Gharbi}, \citenamefont {de~Melo}, \citenamefont
  {Ngo}, \citenamefont {Turmine},\ and\ \citenamefont
  {Vivier}}]{gateman2022use}%
  \BibitemOpen
  \bibfield  {author} {\bibinfo {author} {\bibfnamefont {S.~M.}\ \bibnamefont
  {Gateman}}, \bibinfo {author} {\bibfnamefont {O.}~\bibnamefont {Gharbi}},
  \bibinfo {author} {\bibfnamefont {H.~G.}\ \bibnamefont {de~Melo}}, \bibinfo
  {author} {\bibfnamefont {K.}~\bibnamefont {Ngo}}, \bibinfo {author}
  {\bibfnamefont {M.}~\bibnamefont {Turmine}}, \ and\ \bibinfo {author}
  {\bibfnamefont {V.}~\bibnamefont {Vivier}},\ }\bibfield  {title} {\enquote
  {\bibinfo {title} {On the use of a constant phase element (cpe) in
  electrochemistry},}\ }\href@noop {} {\bibfield  {journal} {\bibinfo
  {journal} {Curr. Opin. Electrochem.}\ }\textbf {\bibinfo {volume} {36}},\
  \bibinfo {pages} {101133} (\bibinfo {year} {2022})}\BibitemShut {NoStop}%
\bibitem [{\citenamefont {Allagui}\ \emph {et~al.}(2020)\citenamefont
  {Allagui}, \citenamefont {Alnaqbi}, \citenamefont {Elwakil}, \citenamefont
  {Said}, \citenamefont {Hachicha}, \citenamefont {Wang},\ and\ \citenamefont
  {Abdelkareem}}]{foedlc}%
  \BibitemOpen
  \bibfield  {author} {\bibinfo {author} {\bibfnamefont {A.}~\bibnamefont
  {Allagui}}, \bibinfo {author} {\bibfnamefont {H.}~\bibnamefont {Alnaqbi}},
  \bibinfo {author} {\bibfnamefont {A.~S.}\ \bibnamefont {Elwakil}}, \bibinfo
  {author} {\bibfnamefont {Z.}~\bibnamefont {Said}}, \bibinfo {author}
  {\bibfnamefont {A.}~\bibnamefont {Hachicha}}, \bibinfo {author}
  {\bibfnamefont {C.}~\bibnamefont {Wang}}, \ and\ \bibinfo {author}
  {\bibfnamefont {M.~A.}\ \bibnamefont {Abdelkareem}},\ }\bibfield  {title}
  {\enquote {\bibinfo {title} {Fractional-order electric double-layer
  capacitors with tunable low-frequency impedance phase angle and energy
  storage capabilities},}\ }\href@noop {} {\bibfield  {journal} {\bibinfo
  {journal} {Appl. Phys. Lett.}\ }\textbf {\bibinfo {volume} {116}},\ \bibinfo
  {pages} {013902} (\bibinfo {year} {2020})}\BibitemShut {NoStop}%
\bibitem [{\citenamefont {Titchmarsh}(1948)}]{titchmarsh1948introduction}%
  \BibitemOpen
  \bibfield  {author} {\bibinfo {author} {\bibfnamefont {E.~C.}\ \bibnamefont
  {Titchmarsh}},\ }\bibfield  {title} {\enquote {\bibinfo {title} {Introduction
  to the theory of fourier integrals},}\ }\href@noop {} {\  (\bibinfo {year}
  {1948})}\BibitemShut {NoStop}%
\bibitem [{\citenamefont {AbdelAty}\ \emph {et~al.}(2018)\citenamefont
  {AbdelAty}, \citenamefont {Elwakil}, \citenamefont {Radwan}, \citenamefont
  {Psychalinos},\ and\ \citenamefont {Maundy}}]{abdelaty2018approximation}%
  \BibitemOpen
  \bibfield  {author} {\bibinfo {author} {\bibfnamefont {A.~M.}\ \bibnamefont
  {AbdelAty}}, \bibinfo {author} {\bibfnamefont {A.~S.}\ \bibnamefont
  {Elwakil}}, \bibinfo {author} {\bibfnamefont {A.~G.}\ \bibnamefont {Radwan}},
  \bibinfo {author} {\bibfnamefont {C.}~\bibnamefont {Psychalinos}}, \ and\
  \bibinfo {author} {\bibfnamefont {B.}~\bibnamefont {Maundy}},\ }\bibfield
  {title} {\enquote {\bibinfo {title} {Approximation of the fractional-order
  laplacian $ s^{\alpha}$ as a weighted sum of first-order high-pass
  filters},}\ }\href@noop {} {\bibfield  {journal} {\bibinfo  {journal} {IEEE
  Trans. Circuits Syst. II Express Briefs}\ }\textbf {\bibinfo {volume} {65}},\
  \bibinfo {pages} {1114--1118} (\bibinfo {year} {2018})}\BibitemShut {NoStop}%
\bibitem [{\citenamefont {Davidson}\ and\ \citenamefont
  {Cole}(1951)}]{davidson1951dielectric}%
  \BibitemOpen
  \bibfield  {author} {\bibinfo {author} {\bibfnamefont {D.~W.}\ \bibnamefont
  {Davidson}}\ and\ \bibinfo {author} {\bibfnamefont {R.~H.}\ \bibnamefont
  {Cole}},\ }\bibfield  {title} {\enquote {\bibinfo {title} {Dielectric
  relaxation in glycerol, propylene glycol, and n-propanol},}\ }\href@noop {}
  {\bibfield  {journal} {\bibinfo  {journal} {J. Chem. Phys.}\ }\textbf
  {\bibinfo {volume} {19}},\ \bibinfo {pages} {1484--1490} (\bibinfo {year}
  {1951})}\BibitemShut {NoStop}%
\bibitem [{\citenamefont {Cole}\ and\ \citenamefont
  {Cole}(1941)}]{cole1941dispersion}%
  \BibitemOpen
  \bibfield  {author} {\bibinfo {author} {\bibfnamefont {K.~S.}\ \bibnamefont
  {Cole}}\ and\ \bibinfo {author} {\bibfnamefont {R.~H.}\ \bibnamefont
  {Cole}},\ }\bibfield  {title} {\enquote {\bibinfo {title} {Dispersion and
  absorption in dielectrics i. alternating current characteristics},}\
  }\href@noop {} {\bibfield  {journal} {\bibinfo  {journal} {J. Chem. Phys.}\
  }\textbf {\bibinfo {volume} {9}},\ \bibinfo {pages} {341--351} (\bibinfo
  {year} {1941})}\BibitemShut {NoStop}%
\bibitem [{\citenamefont {Prabhakar}(1971)}]{prabhakar1971singular}%
  \BibitemOpen
  \bibfield  {author} {\bibinfo {author} {\bibfnamefont {T.~R.}\ \bibnamefont
  {Prabhakar}},\ }\bibfield  {title} {\enquote {\bibinfo {title} {A singular
  integral equation with a generalized mittag leffler function in the
  kernel},}\ }\href@noop {} {\bibfield  {journal} {\bibinfo  {journal}
  {Yokohama Math. J.}\ }\textbf {\bibinfo {volume} {19}},\ \bibinfo {pages}
  {7--15} (\bibinfo {year} {1971})}\BibitemShut {NoStop}%
\bibitem [{\citenamefont {Saxena}, \citenamefont {Mathai},\ and\ \citenamefont
  {Haubold}(2004)}]{saxena2004generalized}%
  \BibitemOpen
  \bibfield  {author} {\bibinfo {author} {\bibfnamefont {R.}~\bibnamefont
  {Saxena}}, \bibinfo {author} {\bibfnamefont {A.}~\bibnamefont {Mathai}}, \
  and\ \bibinfo {author} {\bibfnamefont {H.}~\bibnamefont {Haubold}},\
  }\bibfield  {title} {\enquote {\bibinfo {title} {On generalized fractional
  kinetic equations},}\ }\href@noop {} {\bibfield  {journal} {\bibinfo
  {journal} {Physica A}\ }\textbf {\bibinfo {volume} {344}},\ \bibinfo {pages}
  {657--664} (\bibinfo {year} {2004})}\BibitemShut {NoStop}%
\bibitem [{\citenamefont {Havriliak}\ and\ \citenamefont
  {Negami}(1966)}]{havriliak1966complex}%
  \BibitemOpen
  \bibfield  {author} {\bibinfo {author} {\bibfnamefont {S.}~\bibnamefont
  {Havriliak}}\ and\ \bibinfo {author} {\bibfnamefont {S.}~\bibnamefont
  {Negami}},\ }\bibfield  {title} {\enquote {\bibinfo {title} {A complex plane
  analysis of $\alpha$-dispersions in some polymer systems},}\ }in\ \href@noop
  {} {\emph {\bibinfo {booktitle} {J. Polym. Sci., Part C: Polym. Symp.}}},\
  Vol.~\bibinfo {volume} {14}\ (\bibinfo {organization} {Wiley Online
  Library},\ \bibinfo {year} {1966})\ pp.\ \bibinfo {pages}
  {99--117}\BibitemShut {NoStop}%
\end{thebibliography}
 

 \end{document}